\documentstyle[12pt]{article}

\begin{document}

\title{MULTISPHALERONS IN THE WEINBERG-SALAM THEORY}
\vspace{1.5truecm}
\author{
{\bf Burkhard Kleihaus}\\
Fachbereich Physik, Universit\"at Oldenburg, Postfach 2503\\
D-26111 Oldenburg, Germany
\and
{\bf Jutta Kunz}\\
Fachbereich Physik, Universit\"at Oldenburg, Postfach 2503\\
D-26111 Oldenburg, Germany\\
and\\
Instituut voor Theoretische Fysica, Rijksuniversiteit te Utrecht\\
NL-3508 TA Utrecht, The Netherlands}

\vspace{1.5truecm}

\date{May 26, 1994}

\maketitle
\vspace{1.0truecm}

\begin{abstract}
We construct multisphaleron solutions in the Weinberg-Salam theory.
The multisphaleron solutions carry Chern-Simons charge $n/2$,
where $n$ is an integer,
counting the winding of the fields in the azimuthal angle.
The well-known sphaleron has $n=1$.
The multisphalerons possess axial symmetry
and parity reflection symmetry.
We vary the Higgs mass and the mixing angle.
For small $n$ the energies of the multisphalerons
are on the order of $n$ times the energy of the sphaleron
and their magnetic dipole moments
are on the order of $n$ times
the magnetic dipole moment of the sphaleron.
\end{abstract}
\vfill   \noindent Univ. Utrecht-Preprint THU-94/11
\vfill\eject

\section{Introduction}

In 1976 't Hooft \cite{hooft} observed that
the standard model does not absolutely conserve
baryon and lepton number
due to the Adler-Bell-Jackiw anomaly.
In particular 't Hooft considered spontaneous
fermion number violation due to instanton transitions
between topologically inequivalent vacua.
Recently fermion number violating
tunnelling transitions at high energies
attracted much attention
\cite{ring,ssc}.

In 1983 Manton \cite{man} reconsidered
the possibility of fermion number
violation in the electroweak theory.
By constructing a non-contractible loop
in the configuration space of the Weinberg-Salam theory
Manton predicted the existence of a static, unstable solution
of the field equations,
a sphaleron \cite{km}, representing
the top of the energy barrier between
topologically distinct vacua.

The sphaleron is relevant
in cosmology for the baryon asymmetry of the universe.
At finite temperature the energy barrier between
distinct vacua can be overcome
due to thermal fluctuations of the fields,
and vacuum to vacuum transitions can occur,
accompanied by a change of baryon and lepton number.
The rate for such baryon number violating processes
is largely determined by a Boltzmann factor,
containing the height of the barrier at a given
temperature, and thus by the energy of the sphaleron
\cite{krs,mcl1,mcl2,mcl3,kolb}.

The non-contractible loop constructed by Manton \cite{man}
corresponds to a non-trivial mapping $S_3 \rightarrow S_3$ with
winding number one.
Since $\pi_3(S_3)={\cal Z}$,
there exist non-contractible loops with higher winding numbers.
While the minimal maximal energy configuration along the loops
with winding number one corresponds to the sphaleron,
there should be analogous extremal configurations for the
loops with higher winding numbers.
Correspondingly there should exist further unstable solutions
of the Weinberg-Salam theory, multisphalerons \cite{kk}.

In this paper we investigate the configuration space of
the Weinberg-Salam theory further and construct
multisphaleron solutions \cite{kk}.
In contrast to the sphaleron \cite{km}
which is spherically symmetric for zero mixing angle,
the multisphaleron solutions are only axially symmetric,
even for zero mixing angle \cite{kk}.
The appropriate ansatz for the multisphalerons represents
a generalization of the
axially symmetric ansatz for the sphaleron at
finite mixing angle \cite{kk,kkb1,kkb2},
preserving the invariance under parity.

The ansatz for the multisphalerons contains an integer $n$,
representing the winding number with respect to the azimuthal angle
$\phi$. While $\phi$ covers the full trigonometric circle once,
the fields wind $n$ times around.
The ansatz is thus
analogous to the ansatz of Manton \cite{man2}
and Rebbi and Rossi \cite{rr}
for multimonopole solutions
of an SU(2) gauge theory with a Higgs triplet.
The winding number $n$ also determines the Chern-Simons charge
of the multisphaleron solutions, $N_{\rm CS}=n/2$,
and thus also their baryon number \cite{kk}.

In section 2 we briefly review the Weinberg-Salam lagrangian,
discuss the ansatz and present the resulting energy functional.
We then point out a residual U(1) gauge invariance
of the energy functional.
After fixing the gauge we specify the boundary condition
for the fields.
We present the Chern-Simons charge of the multisphaleron
solutions, subject to these boundary conditions,
and we consider the extraction of their electromagnetic
properties.
In section 3 we discuss the numerical procedure for solving the
seven coupled non-linear partial differential equations.
We then present the numerical results
for the multisphaleron solutions with $n \le 5$
for various values of the Higgs mass.
While the multisphalerons were briefly described in \cite{kk}
for zero mixing angle, we here consider the full range
of the mixing angle, $0 \le \theta_{\rm w} \le \pi/2$.
We present our conclusions in section 4.

\section{\bf Ansatz and energy density}

Let us consider the bosonic sector of the Weinberg-Salam theory
\begin{equation}
{\cal L} = -\frac{1}{4} F_{\mu\nu}^a F^{\mu\nu,a}
- {1 \over 4} f_{\mu \nu} f^{\mu \nu}
+ (D_\mu \Phi)^{\dagger} (D^\mu \Phi)
- \lambda (\Phi^{\dagger} \Phi - \frac{v^2}{2} )^2
\   \end{equation}
with the SU(2)$_{\rm L}$ field strength tensor
\begin{equation}
F_{\mu\nu}^a=\partial_\mu V_\nu^a-\partial_\nu V_\mu^a
            + g \epsilon^{abc} V_\mu^b V_\nu^c
\ , \end{equation}
with the U(1) field strength tensor
\begin{equation}
f_{\mu\nu}=\partial_\mu A_\nu-\partial_\nu A_\mu
\ , \end{equation}
and the covariant derivative for the Higgs field
\begin{equation}
D_{\mu} \Phi = \Bigl(\partial_{\mu}
             -\frac{i}{2}g \tau^a V_{\mu}^a
             -\frac{i}{2}g' A_{\mu} \Bigr)\Phi
\ . \end{equation}
The gauge symmetry is spontaneously broken
due to the non-vanishing vacuum expectation
value $v$ of the Higgs field
\begin{equation}
    \langle \Phi \rangle = \frac{v}{\sqrt2}
    \left( \begin{array}{c} 0\\1  \end{array} \right)
\ , \end{equation}
leading to the boson masses
\begin{equation}
    M_W = \frac{1}{2} g v \ , \ \ \ \
    M_Z = {1\over2} \sqrt{(g^2+g'^2)} v \ , \ \ \ \
    M_H = v \sqrt{2 \lambda}
\ . \end{equation}
The mixing angle $\theta_{\rm w}$ is determined by
the relation $ \tan \theta_{\rm w} = g'/g $,
and the electric charge is $e = g \sin \theta_{\rm w}$.

Gauge field configurations can be classified by a charge,
the Chern-Simons charge.
The Chern-Simons current is not conserved,
its divergence represents the U(1) anomaly of the baryon current.
The SU(2) Chern-Simons current is given by
\begin{equation}
 K_\mu=\frac{g^2}{16\pi^2}\varepsilon_{\mu\nu\rho\sigma} {\rm Tr}(
 {\cal F}^{\nu\rho}
 {\cal V}^\sigma
 + \frac{2}{3} i g {\cal V}^\nu {\cal V}^\rho {\cal V}^\sigma )
\ , \end{equation}
where ${\cal F}_{\nu\rho} = 1/2 \tau^i F^i_{\nu\rho}$, and
${\cal V}_\sigma = 1/2 \tau^i V^i_\sigma$,
and the SU(2) Chern-Simons charge
of a configuration is given by
\begin{equation}
N_{\rm CS} = \int d^3r K^0
\ . \end{equation}
For the vacua the Chern-Simons charge is identical to the
integer winding number,
while the sphaleron has a Chern-Simons charge of 1/2 \cite{km}.

\subsection{\bf Ansatz}

Let us now consider the ansatz for the multisphaleron solutions.
Following Manton \cite{man2} and Rebbi and Rossi \cite{rr} we define
a set of orthonormal vectors
\begin{eqnarray}
\vec u_1^{(n)}(\phi) & = & (\cos n \phi, \sin n \phi, 0) \ ,
\nonumber \\
\vec u_2^{(n)}(\phi) & = & (0, 0, 1) \ ,
\nonumber \\
\vec u_3^{(n)}(\phi) & = & (\sin n \phi, - \cos n \phi, 0)
\ , \end{eqnarray}
and expand the fields as follows
\begin{equation}
V_i^a(\vec r) = u_j^{i(1)}(\phi) u_k^{a(n)}(\phi) w_j^k(\rho,z)
\ , \end{equation}
\begin{equation}
A_i(\vec r) = u_j^{i(1)}(\phi) a_j(\rho,z)
\ , \end{equation}
\begin{equation}
\Phi(\vec r) =i \tau^i u_j^{i(n)}(\phi) h_j(\rho,z)
      \frac{v}{\sqrt2}
    \left( \begin{array}{c} 0\\1  \end{array} \right)
\ . \end{equation}
This ansatz has axial symmetry, i.~e.~a rotation
about the $z$-axis can be compensated by a suitable
gauge transformation \cite{kk,kkb1,kkb2}.
To restrict the ansatz further
we require invariance under the discrete transformation
consisting of charge conjugation and
reflection through the $xz$-plane (mirror symmetry).
This leads to the conditions \cite{kk,kkb1,kkb2}
\begin{equation}
w_1^1(\rho,z)=w_2^1(\rho,z)=w_1^2(\rho,z)=
		   w_2^2(\rho,z)=w_3^3(\rho,z)=0
\ , \end{equation}
\begin{equation}
a_1(\rho,z)=a_2(\rho,z)=0
\ , \end{equation}
\begin{equation}
h_3(\rho,z)=0
\ . \end{equation}

\subsection{\bf Energy density}

The resulting energy functional
\begin{equation}
E = \frac{1}{2} \int ( E_w + E_a + v^2 E_h )
	      \, d\phi \, \rho d\rho \, dz
\   \end{equation}
then has the axially symmetric contributions
\begin{eqnarray}
E_w & = & (\partial_\rho w_3^1 + {1\over{\rho}} ( n w_1^3 + w_3^1 )
	 - g w_1^3 w_3^2 )^2
      +  (\partial_z    w_3^1 + {n\over{\rho}}   w_2^3
	 - g w_2^3 w_3^2 )^2
\nonumber \\
    & + &(\partial_\rho w_3^2 + {1\over{\rho}}   w_3^2
	 + g w_1^3 w_3^1 )^2
      +  (\partial_z    w_3^2
	 + g w_2^3 w_3^1 )^2
      +  (\partial_\rho w_2^3 - \partial_z w_1^3 )^2
\ , \end{eqnarray}
\begin{equation}
E_a = (\partial_\rho a_3 + {1\over{\rho}} a_3 )^2
       + (\partial_z a_3 )^2
\ , \end{equation}
\begin{eqnarray}
E_h & = & (\partial_\rho h_1 - {g\over2} w_1^3 h_2 )^2
       + (\partial_z    h_1 - {g\over2} w_2^3 h_2 )^2
       + (\partial_\rho h_2 + {g\over2} w_1^3 h_1 )^2
       + (\partial_z    h_2 + {g\over2} w_2^3 h_1 )^2
\nonumber \\
     & +& ({n\over{\rho}} h_1
       + {g\over2} ( w_3^1 h_2 - w_3^2 h_1 )
       - {g'\over2} a_3 h_1 )^2
       + (
         {g\over2} ( w_3^1 h_1 + w_3^2 h_2 )
       - {g'\over2} a_3 h_2 )^2
\nonumber \\
     & +& { {\lambda v^2}\over2} ( h_1^2 + h_2^2 - 1 )^2
\ . \end{eqnarray}

\subsection{\bf Residual gauge invariance}

The energy functional
is still invariant under gauge transformations generated by
\begin{equation}
 U= e^{i\Gamma(\rho,z) \tau^i u_3^{i(n)}}
\   \end{equation}
analogous to \cite{kk,kkb1,kkb2,man2,rr}.
Note, that under this transformation the 2-D Higgs doublets
$(h_1,h_2)$ and $(w_3^1,w_3^2-n/g\rho)$ transform with
angle $\Gamma(\rho,z)$ and $2 \Gamma(\rho,z)$, respectively,
while the 2-D gauge field $(w_1^3,w_2^3)$ transforms
inhomogeneously.

Here we fix this gauge degree of freedom by choosing the
`Coulomb gauge' condition \cite{kk,kkb1,kkb2}
\begin{equation}
\partial_\rho w_1^3 + \partial_z w_2^3 =0
\ . \end{equation}
This choice of gauge leads to regular solutions,
while other choices of gauge are known to lead to singular solutions
\cite{kkb2,bkk}.

\subsection{\bf Parity}

In addition to axial, charge conjugation and mirror symmetry
we also require parity reflection symmetry.
Changing to spherical coordinates
and extracting the trivial $\theta$-dependence
we specify the ansatz further \cite{kk,kkb1,kkb2}
\nonumber \\
\begin{eqnarray}
w_1^3(r,\theta) \  &
= &  \ \ {2 \over{gr}} F_1(r,\theta) \cos \theta \ , \ \ \ \
w_2^3(r,\theta) \
= - {2 \over{gr}} F_2(r,\theta) \sin \theta    \ ,
\nonumber \\
w_3^1(r,\theta) \  &
= & - {{2 n}\over{gr}} F_3(r,\theta) \cos \theta    \ , \ \ \ \
w_3^2(r,\theta) \
=  \ \ {{2 n}\over{gr}} F_4(r,\theta) \sin \theta
\ , \end{eqnarray}
\begin{equation}
a_3(r,\theta) = {2 \over{g'r}} F_7(r,\theta) \sin \theta
\ , \end{equation}
\begin{equation}
h_1(r,\theta)   = F_5(r,\theta) \sin \theta \ , \ \ \ \
h_2(r,\theta)   = F_6(r,\theta) \cos \theta
\ . \end{equation}
Note, that the spherically symmetric ansatz for the sphaleron
corresponds to $n=1$ and
$F_1(r,\theta)=F_2(r,\theta)=F_3(r,\theta)=F_4(r,\theta)=f(r)$,
$F_5(r,\theta)=F_6(r,\theta)=h(r)$,
and $F_7(r,\theta)=0$
(where the functions
$f(r)$ and $h(r)$ correspond to those of Ref.~\cite{km}).

\subsection{\bf Boundary conditions}

The above ansatz, when inserted into the classical
equations of motion in the chosen gauge, yields a set of
coupled partial differential equations
for the functions $F_i(r,\theta)$, to be solved numerically
subject to certain boundary conditions.
To obtain regular, finite energy solutions
with the imposed symmetries, we take as
boundary conditions for the functions $F_i(r,\theta)$
\begin{eqnarray}
r=0 & :          & \ \ F_i(r,\theta)|_{r=0}=0,
               \ \ \ \ \ \ i=1,...,7 \ ,
\nonumber \\
r\rightarrow\infty& :
	       & \ \ F_i(r,\theta)|_{r=\infty}=1,
               \ \ \ \ \ i=1,...,6,
               \ \ \ F_7(r,\theta)|_{r=\infty}=0 \ ,
\nonumber \\
\theta=0& :      & \ \ \partial_\theta F_i(r,\theta)|_{\theta=0} =0,
	       \ \ \ i=1,...,7 \ ,
\nonumber \\
\theta=\pi/2& :  & \ \ \partial_\theta F_i(r,\theta)|_{\theta=\pi/2} =0,
               \ i=1,...,7
\ . \end{eqnarray}

\subsection{\bf Chern-Simons charge}

The Chern-Simons charge of the multisphaleron solutions can be evaluated
analogously to the Chern-Simons charge of the sphaleron \cite{km}.
At spatial infinity
the SU(2) vector fields are pure gauge configurations
\begin{equation}
\tau^a V_i^a = -\frac{2i}{g} \partial_i U(\infty) U^\dagger(\infty)
\   \end{equation}
with
\begin{equation}
U(\infty) =  i (\sin \theta \tau^i u_1^{i(n)}
              + \cos \theta \tau^i u_2^{i(n)})
\ . \end{equation}
The proper gauge for evaluating the Chern-Simons charge is
the gauge with $U=1$ at infinity, which is obtained
with the transformation $U^\dagger(\vec r\,)$
\begin{equation}
U(\vec r \,) = \exp \Biggl( i \Omega(r,\theta)
  (\sin \theta \tau^i u_1^{i(n)} + \cos \theta \tau^i u_2^{i(n)}) \Biggr )
\   \end{equation}
and the boundary conditions $\Omega(0) = 0$ and
$\Omega(\infty) = \pi /2$ \cite{km}.
The SU(2) Chern-Simons charge of the multisphaleron solutions
is then given by
\begin{equation}
N_{\rm CS} = {1 \over 2 \pi^2} \int d^3 r Q(\rho,z )
\ , \end{equation}
where the SU(2) Chern-Simons density is determined by \cite{bk}
\begin{eqnarray}
Q(\rho,z) &=&
      n { \sin^2 \Omega \over r^2}
        {\partial \Omega \over \partial r}
\nonumber \\
 &+& n {\partial  \over \partial z}
    \Bigl( { z \over 4 r^3} F_1 \sin (2 \Omega) \Bigr)
  +{n\over \rho} {\partial  \over \partial \rho}
    \Bigl( { \rho ^2 \over 4 r^3} F_2 \sin (2 \Omega) \Bigr)
\nonumber \\
 &+& {\cos^2 \theta \over 4r^2} {\partial \over \partial r}
    \Bigl( F_3 \sin (2 \Omega)  \Bigr)
  +  {\sin^2 \theta \over 4r^2} {\partial \over \partial r}
    \Bigl( F_4 \sin (2 \Omega)  \Bigr)
\nonumber \\
 &+& {1\over 2 \rho} {\partial \over \partial z}
   \Bigl( {{z \rho^2}\over r^3} (F_3 - F_4)
   {\partial \Omega \over \partial \rho} \Bigr)
  -  {1\over 2 \rho} {\partial \over \partial \rho}
   \Bigl( {{z \rho^2}\over r^3} (F_3 - F_4)
   {\partial \Omega \over \partial z} \Bigr)
\ . \end{eqnarray}
Eq.~(30) explicitly demonstrates the effect
of a gauge transformation of the form (28).
Only the first term of $Q(\rho,z)$ determines
the SU(2) Chern-Simons charge,
since the derivative terms do not contribute,
due to the boundary conditions for the functions $F_i(r,\theta)$
and for $\Omega(r,\theta)$.

Since the U(1) field does not
contribute to the Chern-Simons charge \cite{km},
the full Chern-Simons charge is given by the SU(2) Chern-Simons charge.
We thus obtain for the multisphalerons the Chern-Simons charge
\begin{equation}
N_{\rm CS} = n/2
\ , \end{equation}
independently of the Higgs mass and of the mixing angle,
reproducing the well-known Chern-Simons charge of the sphaleron,
$N_{\rm CS} = 1/2$.
The Chern-Simons charge of the sphaleron and of
the multisphalerons corresponds to their
baryonic charge, $Q_B = n/2$ \cite{km}.

\subsection{\bf Magnetic moment}

Another interesting physical quantity characterizing the
electroweak sphaleron is its magnetic moment $\mu$.
The electromagnetic field of the electroweak sphaleron
has the asymptotic behaviour
\begin{equation}
\vec A_{EM}(\vec r) \rightarrow
            {{\vec \mu \times \vec r}\over{4 \pi r^3}}
\ , \end{equation}
where $\vec\mu =(0,0,\mu)$ represents the magnetic moment.
It can be extracted from the long-range behaviour of the functions
$F_3(r,\theta)$, $F_4(r,\theta)$ and $F_7(r,\theta)$.
To extract the magnetic moment we perform
an (asymptotic) gauge transformation, which changes the
asymptotically twisted Higgs field from the `Coulomb gauge'
to the `physical gauge', where the Higgs field is asymptotically
constant.
Applying this transformation to the SU(2) gauge field
yields the `physical' asymptotic isospin-3 component of the
gauge field, needed to construct the asymptotic behaviour of
the electromagnetic field and of the massive $Z^0$-field.
The magnetic moment consists of a sequence of multipoles
\begin{equation}
\mu = \mu_d + \mu_o \frac{5 \cos^2 \theta -1}{r^2} + \ ...
\ , \end{equation}
where the first terms are
the dipole moment $\mu_d$ and the octupole moment
$\mu_o$ \cite{james}.

\subsection{\bf Parameters}

Subject to the boundary conditions (25) we solve the equations
numerically, using the dimensionless coordinate $x=gvr$.
We fix the parameters $g=0.65$ and $M_W=80 {\rm GeV}$.
We vary the Higgs mass, though most calculations presented
are performed for $M_H=M_W$;
and we vary the mixing angle over the full range
$0 \le \theta_{\rm w} \le \pi/2$,
the physical value being $\theta_{\rm w}=0.5$.

\section{\bf Multisphaleron solutions}

To obtain the multisphaleron solutions we start with the $n=1$
sphaleron solution as initial guess and increase the value of $n$
successively.

The numerical calculations are based on the Newton-Raphson
method.
We use the program package FIDISOL \cite{schoen} which also provides
us with an estimate of the discretization errors. The equations are
discretized on a grid covering the integration regions
$0 \leq x \leq 60$ up to $0 \leq x \leq 180$ and
$0 \leq \Theta \leq \pi/2$ with typical grid sizes $50 \times 20$ and
$100 \times 20.$
In most calculations a method with a prescribed grid and order of
consistency is used.
The estimates of the discretization errors are
on the order of $10^{-3}$,
which suffices to extract the physical quantities with high
accuracy.
The calculations are more difficult for large Higgs masses,
$M_H > 100  M_W$.
There a method with optimization of the grid
and of the order of consistency is applied \cite{schoen}.
The accuracy of the solution can be thus improved
until the estimate of the maximal discretization error
is lower than a given limit, e.~g.~$10^{-3}$ in our calculations.

\subsection{\bf Zero mixing angle}

We first present the multisphaleron solutions
in the limit of zero mixing angle.
In this limit the U(1) field
decouples and the function $F_7(x,\theta)$
can consistently be set to zero \cite{kk}.

Let us now consider the gauge invariant properties
of the axially symmetric multisphaleron solutions,
the energy, the energy density
and the length of the Higgs field.
In Fig.~1 we display the energy $E$ of the $n=5$ multisphaleron,
divided by the winding number $n$,
as a function of the Higgs mass and compare it
with the energy of the $n=1$ sphaleron.
We also indicate the energy
of the $n=2,\ 3$ and 4 multisphalerons
at several values of the Higgs mass.
The energy $E$ of the multisphalerons up to $n=5$
is also shown in Table 1 for the Higgs masses
$M_H= M_W/10$, $M_H=M_W$ and $M_H= 10 M_W$
together with the ratio $E/n$.
For small Higgs masses the energy of the multisphalerons is smaller
than $n$ times the energy of the sphaleron.
For large Higgs masses
the energy of the multisphalerons is larger
than $n$ times the energy of the sphaleron.
Inbetween, at a Higgs mass of $M_H\approx M_W$
the energy of the multisphalerons approximately equals
$n$ times the energy of the sphaleron.

In Figs.~2a-2c we show the energy density $\epsilon$,
defined by
\begin{equation}
E= \frac{1}{4\pi} \int \epsilon (\vec x)
                       x^2 dx \sin \theta d\theta d\phi
\ , \end{equation}
where $E$ is the energy in TeV.
The energy density is presented along the $z$-axis,
at $45^\circ$ and along the $\rho$-axis
for the multisphalerons up to $n=5$ and for the Higgs masses
$M_H=M_W/10$, $M_H=M_W$, and $M_H=10 \,M_W$.
The energy density is strongly peaked along the $\rho$-axis,
with the maximum shifting outward with increasing $n$.
Increasing rapidly in size at the origin
for increasing Higgs mass,
the energy density is localized in a decreasing region of space.

In Figs.~3a-3c we show for the corresponding multisphalerons the
magnitude of the Higgs field $\Phi$ in units of $v/\sqrt{2}$,
denoted by the function $L(x,\theta)$
\begin{equation}
L(x,\theta) = \sqrt{ F_5^2(x,\theta) \sin^2 \theta
+ F_6^2(x,\theta) \cos^2 \theta}
\ . \end{equation}
The magnitude of the Higgs field, being zero at the origin,
remains small along the $\rho$-axis for increasingly longer
intervals with increasing $n$.
Like the energy density
the Higgs field is localized in a decreasing region of space
for increasing Higgs mass.

\subsection{\bf Finite mixing angle}

We now consider the case of finite mixing angle
and specify to the Higgs mass $M_H=M_W$.

At finite mixing angle the sphaleron and
the multisphalerons possess a magnetic moment.
The magnetic moment can be extracted from the asymptotic function
$F_7(x,\theta)$ according to \cite{kkb2}
\begin{equation}
\mu = x F_7(x,\theta) {g^2\over{e^2}} \ {e\over{\alpha_{\rm w} M_W}}
\ . \end{equation}
Analyzing the dependence of $\mu$ on the angle $\theta$,
the dipole moment $\mu_d$ and the octupole moment $\mu_o$
can be obtained according to Eq.~(33).
The dipole moment $\mu_d$ can be extracted with good accuracy
up to the mixing angle $\theta_{\rm w}=1.55$,
beyond which the function
$F_7(x,\theta)$ increases rapidly,
leading to large errors on $F_7(x,\theta)$
and making the extraction of $\mu_d$ unreliable.
The octupole moment $\mu_o$ is much smaller
than the dipole moment $\mu_d$,
and can only be extracted with less accuracy.

\subsubsection{\bf Physical mixing angle}

We now consider the multisphalerons at the physical mixing angle
$\theta_{\rm w}=0.5$.

In Table 2 we show the energy (in TeV)
of the multisphalerons up to $n=5$
at the physical mixing angle
and compare it to the energy at zero mixing angle
(for the Higgs mass $M_H=M_W$).
The change in energy due to the finite mixing angle is small,
increasing from 1\% for $n=1$ to 2\% for $n=5$.

In Fig.~4 we show the energy density at the physical mixing angle
along the $z$-axis, at $45^\circ$ and along the $\rho$-axis
for the multisphalerons up to $n=5$.
The change with respect to Fig.~2b, where the energy density
at zero mixing angle is shown, is small.
At the physical mixing angle
the energy density is slightly decreased,
most prominently at the maxima along the $\rho$-axis.

Turning now to the electromagnetic field,
we demonstrate in Fig.~5 that
the magnetic moment consists dominantly
of a dipole moment and an octupole moment.
According to Eq.~(33) the magnetic moment then has a linear dependence
on $\cos^2 \theta$ for a fixed radial coordinate $x$.
{}From the figure the dipole moment and the octupole moment
of the $n=5$ multisphaleron
(for the Higgs mass $M_H=M_W$) can be extracted.

Table 2 gives in addition to the energy
the dipole moment $\mu_d$ and the octupole moment $\mu_o$
for the multisphalerons up to $n=5$
at the physical mixing angle $\theta_{\rm w}=0.5$
(for the Higgs mass $M_H=M_W$).
The octupole moment of the sphaleron is very small \cite{james}.
It was therefore not extracted in our previous calculations
of the sphaleron at finite mixing angle, from where the data
for the table are taken \cite{kkb1,kkb2}.
A perturbative calculation of the octupole moment of the sphaleron
at the physical mixing angle
lead to $\mu_o \approx 10^{-1} e/(\alpha_{\rm w} M_W^3)$
\cite{james}.

The functions $F_i(x,\theta)$ are gauge dependent quantities.
Nevertheless we show as an example
the functions $F_i(x,\theta)$ for the $n=5$ multisphaleron
at the physical mixing angle $\theta_{\rm w}=0.5$
(for the Higgs mass $M_H=M_W$)
in Figs.~6a-6d.

\subsubsection{\bf Mixing angle dependence}

We now demonstrate the dependence of the multisphalerons on
the mixing angle over the full range of the mixing angle
$0 \le \theta_{\rm w} \le \pi/2$.

In Fig.~7 we show the energy of the $n=5$ multisphaleron
as a function of the mixing angle
and the energy of the $n=2,\ 3$ and 4 multisphalerons
at the mixing angles $\theta_{\rm w}=0$, 0.5 and 1.0.
The figure also gives the energy of the sphaleron
for comparison \cite{kkb2}.
With increasing mixing angle $\theta_{\rm w}$
the energy decreases monotonically.
Approaching the limiting value $\theta_{\rm w} = \pi/2$
the energy of the $n=5$ multisphaleron smoothly approaches
a limiting value. This is in contrast to the sphaleron,
whose energy decreases strongly, when
$\theta_{\rm w} \rightarrow \pi/2$.
(Though recall, that the numerics becomes less reliable
in the limit $\theta_{\rm w} \rightarrow \pi/2$.)

The energy density is presented in Figs.~8a-8b.
In Fig.~8a we show the energy density for the
multisphalerons up to $n=5$ at the mixing angle
$\theta_{\rm w}=1$.
Comparing this figure with Fig.~4
we observe a further decrease of the energy density.
In Fig.~8b we show the energy density
of the $n=5$ multisphaleron for
the mixing angles $\theta_{\rm w}=$0.5, 1.0, 1.5 and 1.55.
While the energy density decreases monotonically with
increasing mixing angle, the maximum along the $\rho$-axis
remains almost fixed.

In Fig.~9 we present the dipole moment as a function of
the mixing angle for the $n=5$ multisphaleron.
We also show the dipole moment
of the $n=2,\ 3$ and 4 multisphalerons
at the mixing angles $\theta_{\rm w}=0.5$ and 1.0.
For comparison the dipole moment of the sphaleron
is also shown \cite{kkb2}.
We observe, that the dipole moment of the multisphalerons
is on the order of $n$ times the dipole moment of the sphaleron,
for all mixing angles.

Table 3 summarizes the energy, the dipole moment and the octupole moment
of the $n=2$, 3, 4 and 5 multisphalerons
at the mixing angle $\theta_{\rm w}=1.0$.

\section{\bf Conclusions}

We have constructed multisphaleron solutions
in the electroweak interactions.
The multisphaleron solutions have axial symmetry
and parity reflection symmetry.
At zero mixing angle,
where the sphaleron is spherically symmetric,
the multisphalerons are only axially symmetric, too.

The multisphaleron solutions are characterized by an integer $n$,
related to the winding of the fields in the azimuthal angle $\phi$.
With increasing $n$ the energy density of the
multisphalerons becomes increasingly deformed,
the maximum of the energy density occurring along the $\rho$-axis
at increasing distance from the origin.

The energy of the multisphalerons
is on the order of $n$ times the sphaleron energy.
For small Higgs masses the energy
of the multisphalerons is less than $n$ times
the energy of the sphaleron, $E(n)< nE(1)$,
while for large Higgs masses the energy
of the multisphalerons is greater than $n$ times
the energy of the sphaleron, $E(n)> nE(1)$.
Inbetween, at a Higgs mass of $M_H\approx M_W$
the energy of the multisphalerons approximately equals
$n$ times the energy of the sphaleron.

At the physical mixing angle the multisphaleron solutions differ
only little from the solutions at zero mixing angle,
except that the U(1) field is finite.
Compared to zero mixing angle
the energy is only smaller by 1\% for the sphaleron
and by 2\% for the $n=5$ multisphaleron
at the physical mixing angle.
Like the sphaleron the multisphalerons have large magnetic
dipole moments, being on the order of $n$ times the dipole moment
of the sphaleron. Their octupole moments are small and negative.

The sphaleron represents the maximal energy configuration
along a non-contractible loop with winding number $n=1$,
its Chern-Simons charge is given by $N_{\rm CS}=1/2$.
The multisphaleron solutions have
Chern-Simons charge $N_{\rm CS}=n/2$.
We conclude, that symmetric
non-contractible loops can be constructed,
leading from the vacuum with Chern-Simons charge $N_{\rm CS}=0$
to topologically distinct vacua with
Chern-Simons charge $N_{\rm CS}=n$,
passing the multisphaleron solutions with $N_{\rm CS}=n/2$
midway.

At finite temperature such paths involving multisphalerons
should allow for thermal
fermion number violating transitions,
which would correspond to
tunnelling transitions via multiinstantons
at zero temperature.
Since multiinstantons with winding number $n$
possess $n$ fermion zero modes,
we expect to encounter $n$ fermion zero modes
along vacuum to vacuum paths, passing the $n$-th multisphaleron.
Though relevant in principle for baryon number violating processes,
the multisphalerons have much higher energies
than the sphaleron, making such processes much more
unlikely and therefore less relevant for cosmology and
for the baryon asymmetry of the universe.

{\bf \sl Acknowledgement}

We gratefully acknowledge discussions with Yves Brihaye.

\vfill\eject

\vfill\eject
\section{\bf Tables}
\begin{center}
{\bf Table 1} \\
\vspace{3 mm}
$E(n)$ [TeV] ($E(n)/n$ [TeV])\\
\vspace{3 mm}
\begin{tabular}{||c|c|c|c||} \hline\hline
$n$   & $M_H =  M_W/10 $ & $ M_H = M_W$ & $M_H = 10 M_W $\\
\hline
$ 1 $ & $ 7.468$ $(7.468)$ & $ 8.665$ $(8.665)$  & $11.375$ $(11.375)$ \\
$ 2 $ & $13.791$ $(6.896)$ & $17.140$ $(8.570)$  & $24.937$ $(12.469)$ \\
$ 3 $ & $19.881$ $(6.627)$ & $25.879$ $(8.626)$  & $40.030$ $(13.343)$ \\
$ 4 $ & $25.908$ $(6.477)$ & $34.931$ $(8.733)$  & $56.298$ $(14.075)$ \\
$ 5 $ & $31.931$ $(6.386)$ & $44.288$ $(8.858)$  & $73.521$ $(14.704)$ \\
\hline
\hline
\end{tabular}
\end{center}

{\bf Table 1:}

The energy $E$ (in TeV)
and the energy $E$ (in TeV) divided by the winding number $n$
are shown for the sphaleron and the multisphalerons up to $n=5$
for the Higgs masses $M_H=M_W/10$, $M_H=M_W$ and $M_H=10 M_W$
for zero mixing angle.

\begin{center}
{\bf Table 2} \\
\vspace{3 mm}
$\theta_{\rm w}=0.5$, $M_H=M_W$\\
\vspace{3 mm}
\begin{tabular}{||c|c|c|c|c||} \hline\hline
$n$   & $E$ [TeV] & $E(0.5)/E(0)$ & $\mu_d\ [e/(\alpha_{\rm w} M_W)]$ &
$\mu_o\ [e/(\alpha_{\rm w} M_W^3)]$ \\
\hline
$ 1 $ & $ 8.587$ & $0.991$ & $ 1.779$ & $    $ \\
$ 2 $ & $16.923$ & $0.987$ & $ 3.725$ & $ -2.$ \\
$ 3 $ & $25.475$ & $0.984$ & $ 5.947$ & $ -7.$ \\
$ 4 $ & $34.294$ & $0.982$ & $ 8.420$ & $-14.$ \\
$ 5 $ & $43.374$ & $0.979$ & $11.116$ & $-24.$ \\
\hline
\hline
\end{tabular}
\end{center}

{\bf Table 2:}

The energy $E$ (in TeV),
the ratio of the energy
at the physical mixing angle $\theta_{\rm w}=0.5$
to the energy at zero mixing angle $\theta_{\rm w}=0$,
the magnetic dipole moment $\mu_d$
(in units of $[e/(\alpha_{\rm w} M_W)]$)
and the magnetic octupole moment $\mu_o$
(in units of $[e/(\alpha_{\rm w} M_W^3)]$)
are shown for the sphaleron and the multisphalerons up to $n=5$
for the Higgs mass $M_H=M_W$
and for the physical mixing angle $\theta_{\rm w}=0.5$.
(The magnetic octupole moment $\mu_o$ was not extracted in
\cite{kkb2}.)

\newpage
\begin{center}
{\bf Table 3} \\
\vspace{3 mm}
$\theta_{\rm w}=1.0$, $M_H=M_W$\\
\vspace{3 mm}
\begin{tabular}{||c|c|c|c|c||} \hline\hline
$n$   & $E$ [TeV] & $E(1.0)/E(0)$ & $\mu_d\ [e/(\alpha_{\rm w} M_W)]$ &
$\mu_o\ [e/(\alpha_{\rm w} M_W^3)]$ \\
\hline
$ 1 $ & $ 8.295$ & $0.957$ & $ 1.848$ & $    $ \\
$ 2 $ & $16.179$ & $0.944$ & $ 3.835$ & $ -3.$ \\
$ 3 $ & $24.169$ & $0.934$ & $ 6.051$ & $ -7.$ \\
$ 4 $ & $32.327$ & $0.925$ & $ 8.490$ & $-15.$ \\
$ 5 $ & $40.659$ & $0.918$ & $11.116$ & $-25.$ \\
\hline
\hline
\end{tabular}
\end{center}

{\bf Table 3:}

Idem Table 2 for the mixing angle $\theta_{\rm w}=1.0$.

\vfill\eject

\section{\bf Figure captions}

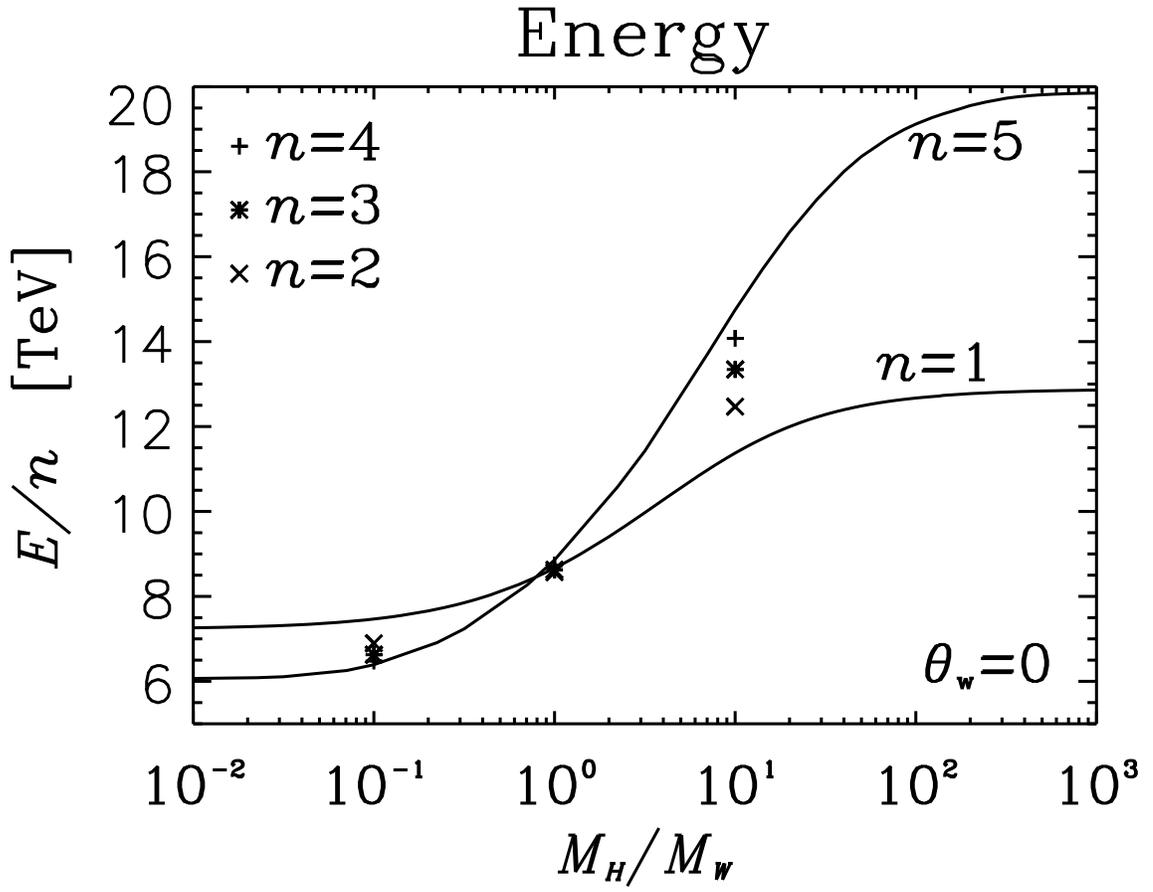
{\bf Figure 1:}

The energy $E$ (in TeV), divided by the winding number $n$,
is shown as a function of the Higgs mass
$M_H$ (in units of $M_W$)
for zero mixing angle
for the sphaleron and the $n=5$ multisphaleron (solid lines),
the $n=2$ multisphaleron (crosses),
the $n=3$ multisphaleron (asterisks),
and the $n=4$ multisphaleron (plus signs).

{\bf Figure 2a:}

The energy density $\epsilon(x,\theta)$ (in TeV) is shown
as a function of the dimensionless coordinate $x$
for the Higgs mass $M_H=M_W/10$
and for zero mixing angle
for the sphaleron and the multisphalerons up to $n=5$
along the $z$-axis at $\theta=0^\circ$ (dashed),
at $\theta=45^\circ$ (dotted),
and along the $\rho$-axis at $\theta=90^\circ$ (solid).

{\bf Figure 2b:}

Idem Fig.~2a for the Higgs mass $M_H=M_W$.

{\bf Figure 2c:}

Idem Fig.~2a for the Higgs mass $M_H=10M_W$.
The density profile along the $z$-axis and at $\theta=45^\circ$
becomes steeper with increasing winding number $n$.

{\bf Figure 3a:}

The magnitude of the Higgs field $L(x,\theta)$ (in units of $v/\sqrt{2}$)
is shown as a function of the dimensionless coordinate $x$
for the Higgs mass $M_H=M_W/10$
and for zero mixing angle
for the sphaleron and the multisphalerons up to $n=5$
along the $z$-axis at $\theta=0^\circ$ (dashed),
and along the $\rho$-axis at $\theta=90^\circ$ (solid).
Both, along the $z$-axis and along the $\rho$-axis,
the curves become flatter with increasing winding number $n$.

{\bf Figure 3b:}

Idem Fig.~3a for the Higgs mass $M_H=M_W$.

{\bf Figure 3c:}

Idem Fig.~3a for the Higgs mass $M_H=10M_W$.
But with increasing winding number $n$,
the curves become steeper along the $z$-axis,
while they become flatter along the $\rho$-axis.

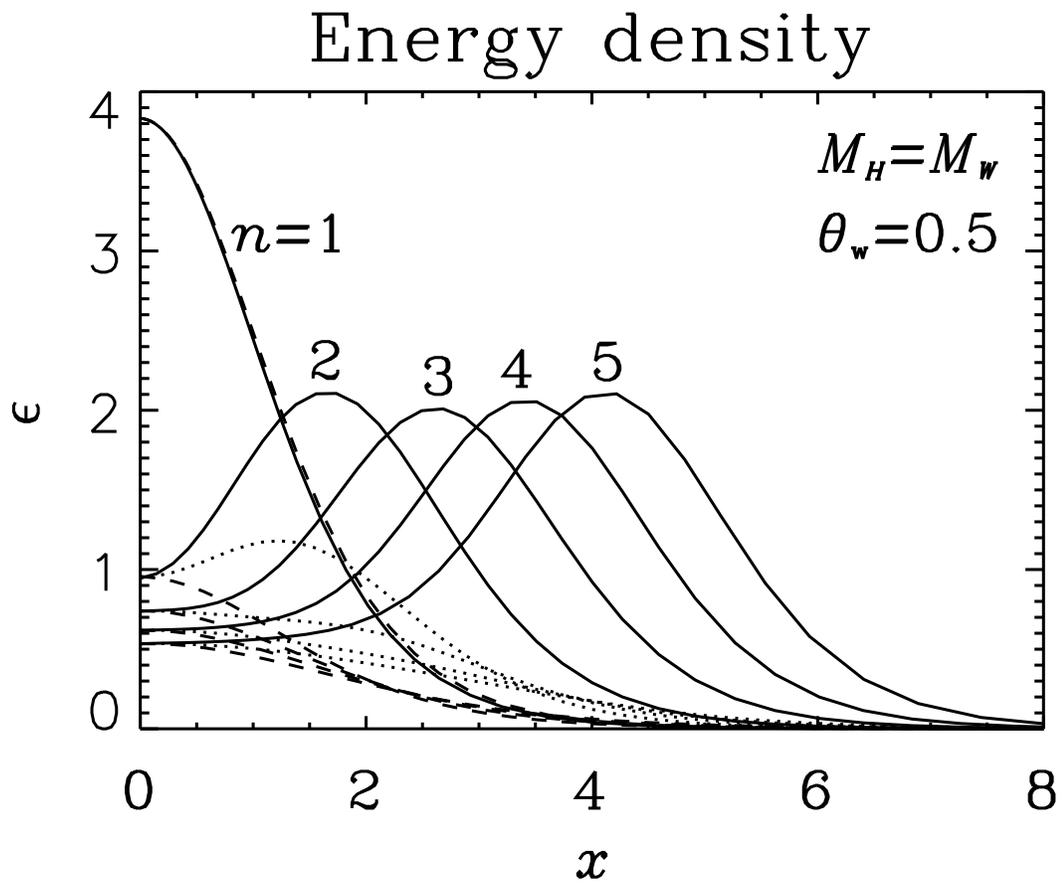
{\bf Figure 4:}

The energy density $\epsilon(x,\theta)$ (in TeV) is shown
as a function of the dimensionless coordinate $x$
for the Higgs mass $M_H=M_W$
and for the physical mixing angle $\theta_{\rm w}=0.5$
for the sphaleron and the multisphalerons up to $n=5$
along the $z$-axis at $\theta=0^\circ$ (dashed),
at $\theta=45^\circ$ (dotted),
and along the $\rho$-axis at $\theta=90^\circ$ (solid).

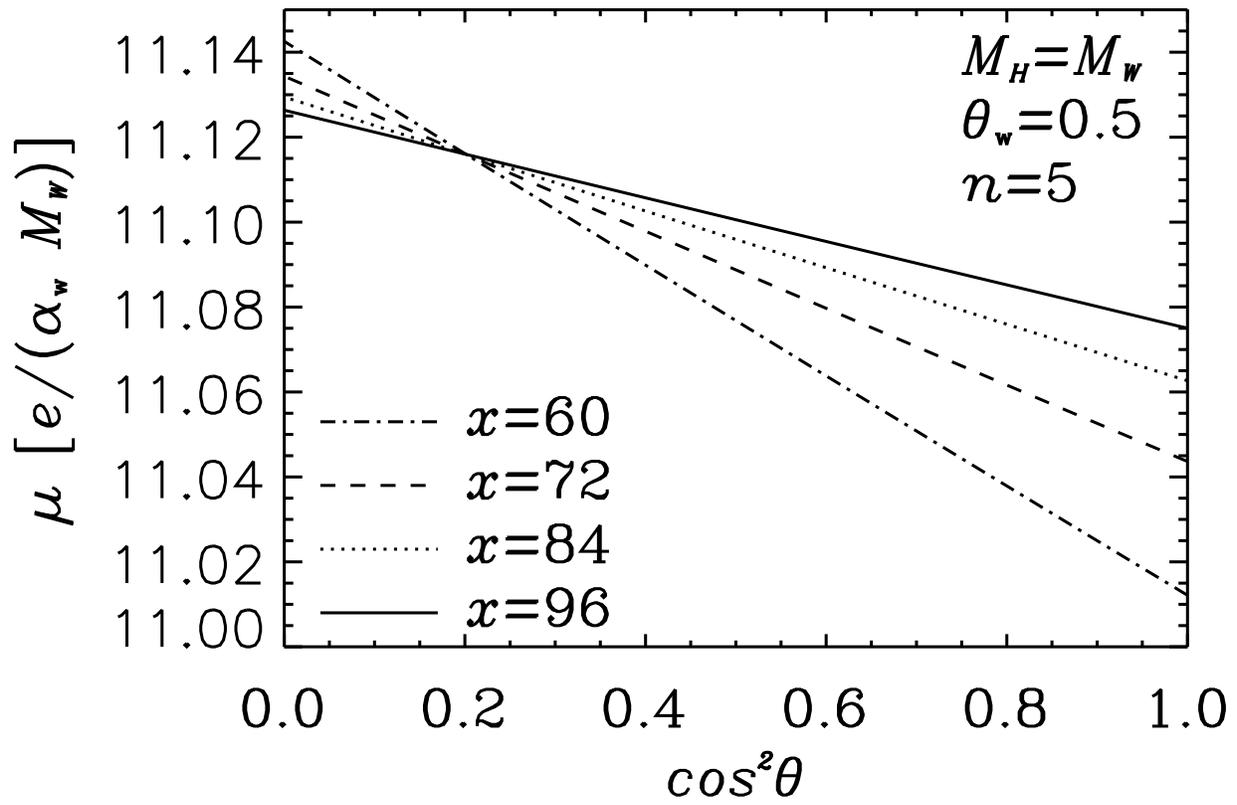
{\bf Figure 5:}

The magnetic moment $\mu$
(in units of $e/(\alpha_{\rm w} M_W$)),
is shown as a function of $\cos^2 \theta$
for the Higgs mass $M_H=M_W$
and for the physical mixing angle $\theta_{\rm w}=0.5$
for the $n=5$ multisphaleron
for the values of the dimensionless coordinate
$x=60$ (dash-dotted), $x=72$ (dashed), $x=84$ (dotted),
and $x=96$ (solid).

\newpage
{\bf Figure 6a:}

The SU(2) gauge field functions
$F_1(x,\theta)$ and $F_2(x,\theta)$ are shown
as a function of the dimensionless coordinate $x$
for the Higgs mass $M_H=M_W$
and for the physical mixing angle $\theta_{\rm w}=0.5$
for the $n=5$ multisphaleron
along the $z$-axis at $\theta=0^\circ$ (dashed),
at $\theta=45^\circ$ (dotted),
and along the $\rho$-axis at $\theta=90^\circ$ (solid).

{\bf Figure 6b:}

Idem Fig.~6a for the SU(2) gauge field functions
$F_3(x,\theta)$ and $F_4(x,\theta)$.

{\bf Figure 6c:}

Idem Fig.~6a for the Higgs field functions
$F_5(x,\theta)$ and $F_6(x,\theta)$.

{\bf Figure 6d:}

Idem Fig.~6a for the magnitude of the Higgs field $L(x,\theta)$
and the U(1) gauge field function $F_7(x,\theta)$.

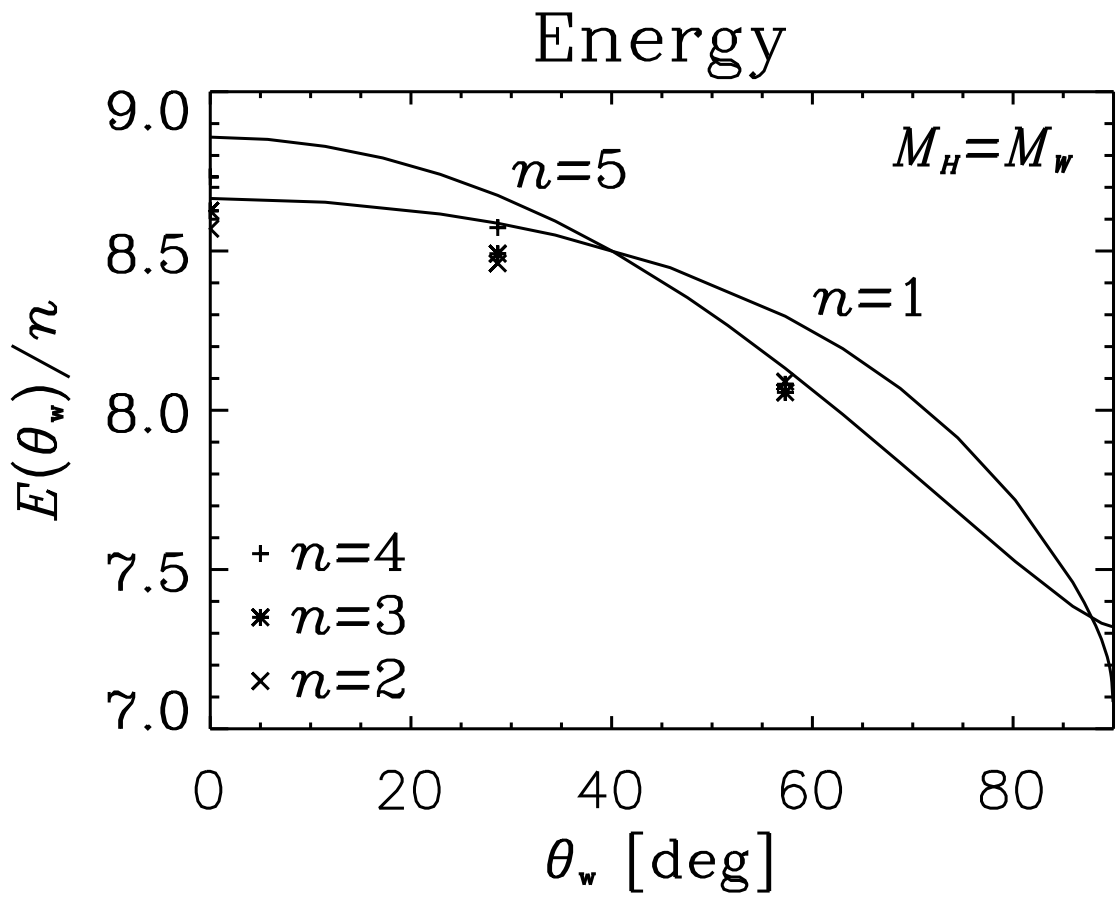
{\bf Figure 7:}

The energy $E$ (in TeV), divided by the winding number $n$,
is shown as a function of the mixing angle $\theta_{\rm w}$
for the Higgs mass $M_H=M_W$
for the sphaleron and the $n=5$ multisphaleron (solid lines),
the $n=2$ multisphaleron (crosses),
the $n=3$ multisphaleron (asterisks),
and the $n=4$ multisphaleron (plus signs).

{\bf Figure 8a:}

The energy density $\epsilon(x,\theta)$ (in TeV) is shown
as a function of the dimensionless coordinate $x$
for the Higgs mass $M_H=M_W$
and for the mixing angle $\theta_{\rm w}=1.0$
for the sphaleron and the multisphalerons up to $n=5$
along the $z$-axis at $\theta=0^\circ$ (dashed),
at $\theta=45^\circ$ (dotted),
and along the $\rho$-axis at $\theta=90^\circ$ (solid).

{\bf Figure 8b:}

The energy density $\epsilon(x,\theta)$ (in TeV) is shown
as a function of the dimensionless coordinate $x$
for the Higgs mass $M_H=M_W$
for the $n=5$ multisphaleron
for the mixing angles $\theta_{\rm w}=0.5$ (solid),
$\theta_{\rm w}=1.0$ (dot-dashed),
$\theta_{\rm w}=1.5$ (dotted),
and $\theta_{\rm w}=1.55$ (dashed),
along the $z$-axis at $\theta=0^\circ$,
at $\theta=45^\circ$,
and along the $\rho$-axis at $\theta=90^\circ$,
$\epsilon(x,\theta)$ is
monotonically increasing with increasing angle.

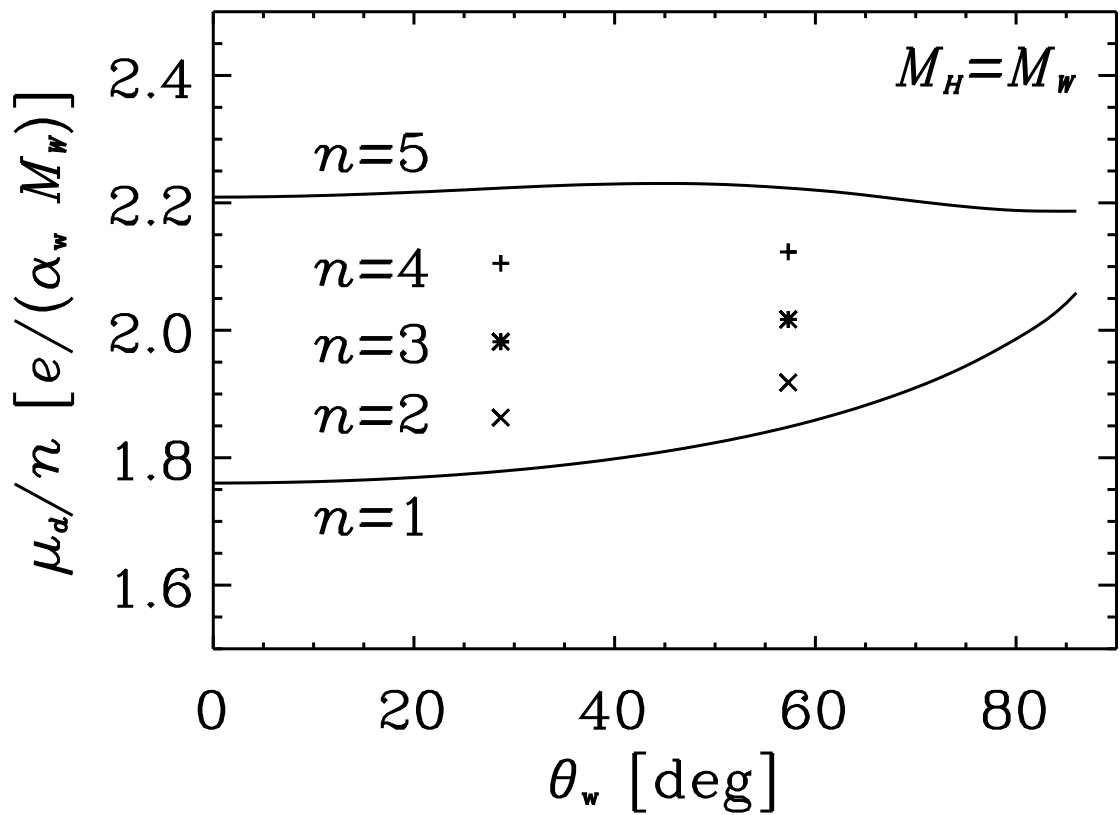
{\bf Figure 9:}

The magnetic dipole moment $\mu_d$
(in units of $e/(\alpha_{\rm w} M_W$)),
divided by the winding number $n$,
is shown as a function of the mixing angle $\theta_{\rm w}$
for the Higgs mass $M_H=M_W$
for the sphaleron and the $n=5$ multisphaleron (solid lines),
the $n=2$ multisphaleron (crosses),
the $n=3$ multisphaleron (asterisks),
and the $n=4$ multisphaleron (plus signs).

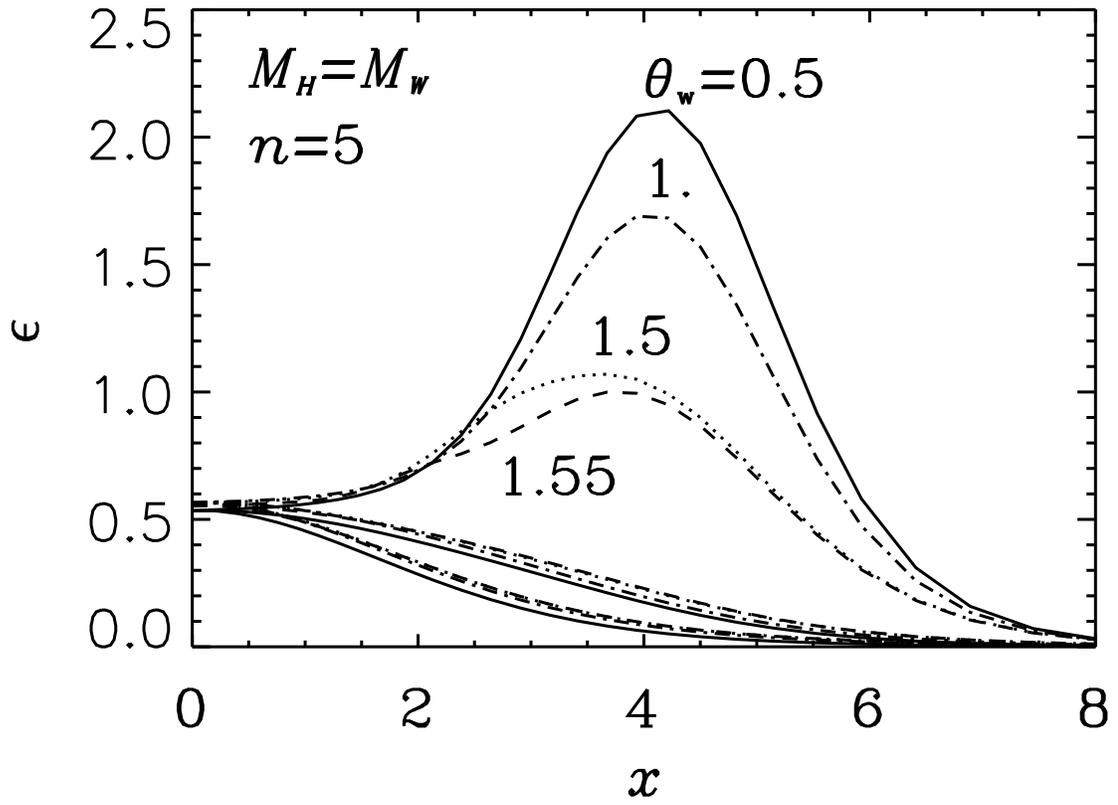
\end{document}